\documentclass[runningheads]{llncs}

\usepackage[]{graphicx}
\usepackage{subfigure}
\usepackage[T1]{fontenc}
\usepackage[utf8]{inputenc}

\usepackage{amsmath}
\usepackage{algpseudocode,algorithm,algorithmicx}

\usepackage{booktabs}
\usepackage{multirow}
\usepackage{tabularx}

\title{Efficient Discovering of Top-K Sequential Patterns in Event-Based Spatio-Temporal Data}

\titlerunning{Top-K Sequential Patterns in Spatio-Temporal Data}

\author{Piotr S. Maci\k{a}g} \institute{Institute of Computer Science, Warsaw University of Technology\\
		Nowowiejska 15/19,\\
		00-665, Warsaw, Poland,\\
		email: pmaciag@ii.pw.edu.pl}

\begin{document}
	
\maketitle
	
\begin{abstract}
	We consider the problem of discovering sequential patterns from event-based spatio-temporal data. The dataset is described by a set of event types and their instances. Based on the given dataset, the task is to discover all significant sequential patterns denoting some attraction relation between event types occurring in a pattern. Already proposed algorithms discover all significant sequential patterns based on the significance threshold, which minimal value is given by an expert. Due to the nature of described data and complexity of discovered patterns, it may be very difficult to provide reasonable value of significance threshold. We consider the problem of effective discovering of $K$ most important patterns in a given dataset (that is discovering of Top-$K$ patterns).
\end{abstract}

\section{Introduction}

\begin{figure}[h!t]
	\centering 
	\includegraphics[width=0.8\linewidth]{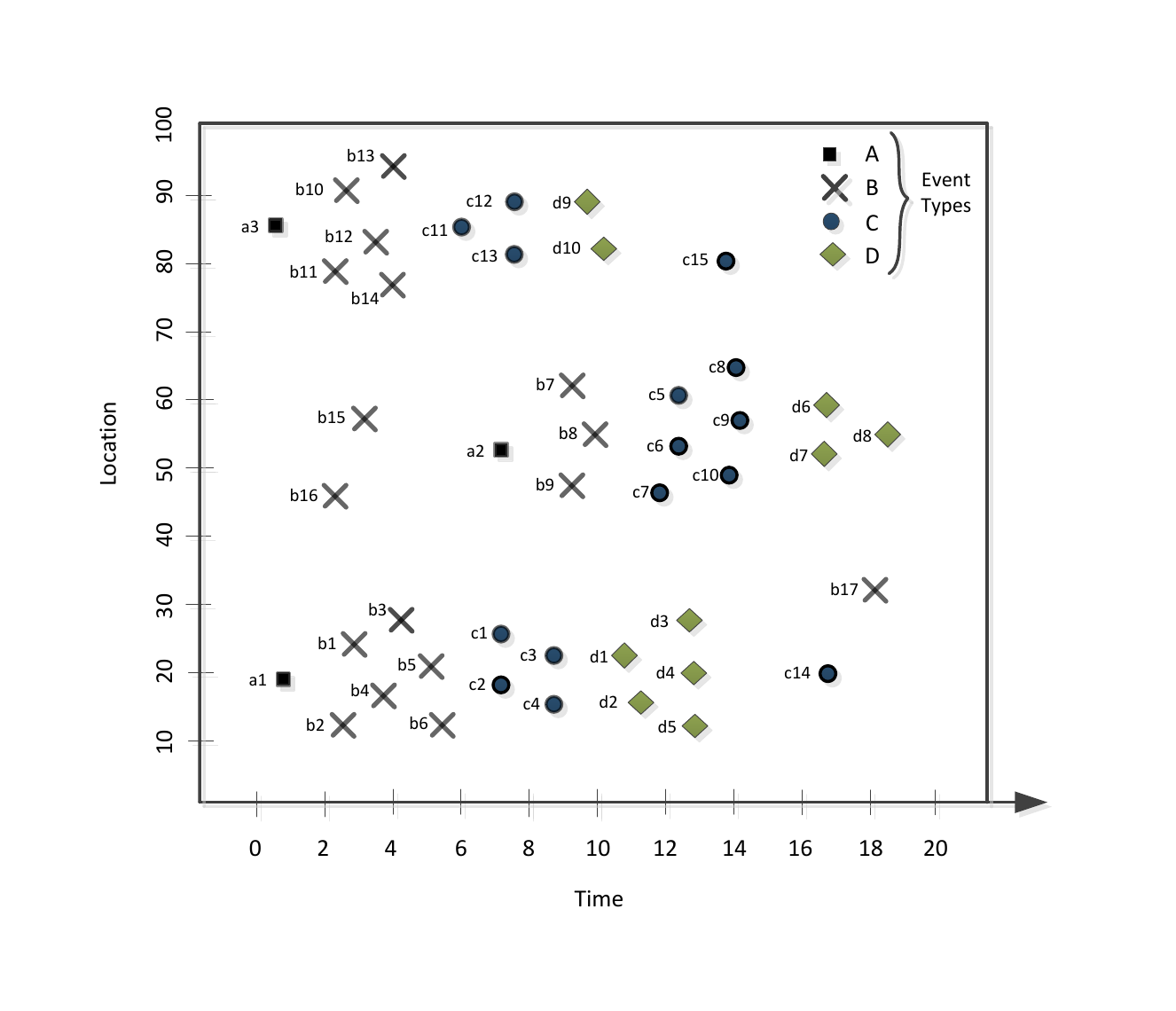}
	\caption{An example of the spatio-temporal event-based dataset from Table \ref{Table:ExamDataset}.}
	\label{Fig:1}      
\end{figure}

Discovering knowledge from spatio-temporal data is gaining attention of researchers nowadays. Based on literature we can distinguish two basic types of spatio-temporal data: event-based and trajectory-based \cite{ref1284:Li2014}. Event-based spatio-temporal data is described by a set of event types $F = \{f_1, f_2, $ $\dots, f_n\}$ and a set of instances $D$. Each instance $e \in D$ denotes an occurrence of a particular event type from $F$ and is associated with instance identifier, location in spatial dimension and occurrence time. Figure \ref{Fig:1} provides possible sets $D = \{a1, a2, \dots, d10\}$ and $F = \{A, B, C, D\}$. Event-based spatio-temporal data and the problem of discovering frequent sequential patterns in this type of data have been introduced in \cite{ref1284:Huang2008}.

\begin{table}
	\centering
	\caption{An example of a spatio-temporal event-based dataset.}
	\label{Table:ExamDataset}       
	\begin{tabular}{llll}
		\hline\noalign{\smallskip}
		Instance identifier & Event type & Spatial location & Occurrence time  \\
		\noalign{\smallskip}\hline\noalign{\smallskip}
		a1 & A & 19 & 1 \\
		a2 & A & 83 & 1 \\
		\vdots & \vdots & \vdots & \vdots \\
		b1 & B & 25 & 3 \\
		b2 & B & 1 & 3 \\
		\vdots & \vdots & \vdots & \vdots \\
		c1 & C & 25 & 7 \\
		c2 & C & 15 & 7 \\
		\vdots & \vdots & \vdots & \vdots \\
		d1 & D & 21 & 11 \\
		d2 & D & 13 & 12 \\
		\vdots & \vdots &  \vdots & \vdots \\
		\noalign{\smallskip}\hline
	\end{tabular}
\end{table}

The task of mining spatio-temporal sequential patterns in given datasets $F$ and $D$ may be defined as follows. We assume that the \textit{following} relation (or attraction relation) $f_{1} \rightarrow f_2$ between any two event types in $F$ denotes the fact, that instances of type $f_1$ attract in their spatial and temporal neighborhoods occurrences of instances of type $f_2$. The strength of the following relation $f_1 \rightarrow f_2$ is investigated by comparing the density of instances of type $f_2$ in spatio-temporal neighborhoods of instances of type $f_1$ and density of instances of type $f_2$ in the whole spatio-temporal embedding space $V$. We provide the strict definition of density in Section \ref{Sec:Basic notions}. The problem introduced in \cite{ref1284:Huang2008} is to discover all significant sequential patterns defined in the form $f_1 \rightarrow f_2 \rightarrow \dots \rightarrow f_k$, where the significance threshold is given by an expert. In contrary to this approach, we consider the problem of discovering $K$ most significant patterns in the given dataset. Providing significance threshold for discovering patterns may be difficult due to the complex nature of considered task. 

The rest of the paper is organized as follows. In Section \ref{Sec:Basic notions}  we provide elementary notions. Section \ref{Sec:TopKPatterns} gives our method and main results. Section \ref{sec:Exp} provides basic experiments. In Section \ref{Sec:Conclusions} we give conclusions and future problems. The main results of the paper are: introduction of a notion of top-K patterns in event-based spatio-temporal data, analysis and definition of the algorithm discovering top-K patterns and experimental results showing efficiency of proposed approach.

\section{Basic notions}
\label{Sec:Basic notions}

The dataset given in Fig.~\ref{Fig:1} is contained in the spatiotemporal space $V$, which temporal dimension is of size $20$ and spatial location is provided by numbers between $0$ and $100$. For simplicity we denote spatial location in only one dimension. Usually, spatial location is defined by two dimensions (f.e. geographical coordinates). By $|V|$ we denote the volume of space $V$, calculated as the product of spatial area and size of time dimension. Spatial and temporal sizes of spatiotemporal space are usually given by an expert. For example, for Fig.~\ref{Fig:1} $|V| = 20 * 100 = 2000$. In the following definitions and notions we use terms sequential patterns and sequence interchangeably.

\begin{definition}{Neighborhood space.}
	$V_{N(e)}$ we denote the neighborhood space of instance $e$. The shape of $V_{N(e)}$ is given by an expert. If $V_{N(e)}$ has cylindrical or conical shape, then $\mathcal{R}$ denotes the spatial radius and $\mathcal{T}$ temporal interval of that space.  If $V_{N(e)}$ has cubic shape, then $\mathcal{R}$ may denote size of spatial square and $\mathcal{T}$ temporal interval of that space. Consider example given in Fig.\ref{Fig:1} where we denote neighborhood spaces $V_{N(a1)}, V_{N(a2)}, V_{N(a3)}$.
	\label{Def:NeighborhoodSpace}
\end{definition}

\begin{definition}{Neighborhood definition \cite{ref1284:Huang2008}.}
	For a given event type $f$ and an occurrence of event instance $e$ of that type, the neighborhood of $e$ is defined as follows:
	\begin{equation}
	\begin{split}
	N(e) = & \{p | p \in D \land distance(e.location, p.location) \leq \mathcal{R} \\ 
	& \land (p.time - e.time) \leq \mathcal{T}\}
	\end{split}
	\end{equation}
	where $\mathcal{R}$ denotes the spatial radius and $\mathcal{T}$ temporal interval of the neighborhood space $V_{N(e)}$.
	\label{Def:Neighborhood}
\end{definition}

\begin{definition}{Density \cite{ref1284:Huang2008}.}
	For a given spatiotemporal space $V$, event type $f$ and its events instances in $D$, density is defined as follows:
	\begin{equation}
	\begin{split}
	Density(f, V) = \frac{|\{e | e \in f \land \text{e is inside V}\}|}{|V|}
	\end{split}
	\end{equation}
	that is, density is the number of instances of type $f$ occurring inside some space $V$ divided by the volume of that space.
	\label{Def:Density}
\end{definition}

\begin{definition}{Density ratio \cite{ref1284:Huang2008}.}
	Density ratio for two event types $f_1, f_2$ and their instances is defined as follows:
	\begin{equation}
	\begin{split}
	DensityRatio(f_1 \rightarrow f_2) = \frac{avg_{e \in f_1}(Density(f_2, V_{N(e)}))}{Density(f_2, V)}
	\end{split}
	\end{equation}
	where $\rightarrow$ denotes the \textit{following} relation between event types $f_1, f_2$. \\ 
	$avg_{e \in f_1}(Density(f_2, V_{N(e)}))$ specifies the average density of instances of type $f_2$ occurring in the neighborhood spaces $V_{N(e)}$ defined for instances $e \in f_1$. $V$ denotes the whole considered spatiotemporal space and $Density(f_2, V)$ specifies density of instances of type $f_2$ in that space.
	\label{Def:DensityRatio}
\end{definition}

\begin{definition}{Sequence $\overrightarrow{s}$ and tailEventSet($\overrightarrow{s}$) \cite{ref1284:Huang2008}.}
	$\overrightarrow{s}$ denotes a k-length sequence of event types: $s[1] \rightarrow s[2] \rightarrow \dots \rightarrow s[k-1] \rightarrow s[k]$. tailEventSet($\overrightarrow{s}$) denotes the set of instances of type $\overrightarrow{s}[k]$ participating in the sequence $\overrightarrow{s}$.
	\label{Def:Sequence}
\end{definition}

\begin{definition}{Sequence index \cite{ref1284:Huang2008}.}
	For a given k-length sequence $\overrightarrow{s}$, sequence index is defined as follows:
	\begin{enumerate}
		\item When $k = 2$ then:
		\begin{equation} 
		SeqIndex(\overrightarrow{s}) = DensityRatio(\overrightarrow{s}[1] \rightarrow \overrightarrow{s}[2])
		\end{equation}
		\item When $k > 2$ then:
		\begin{equation} 			
		SeqIndex(\overrightarrow{s}) = \text{min} \left \{
		\begin{array}{ll}
		SeqIndex(\overrightarrow{s}[1:k-1]),  \\
		DensityRatio(\overrightarrow{s}[k-1] \rightarrow \overrightarrow{s}[k])
		\end{array}
		\right.
		\end{equation}
	\end{enumerate}
	\label{Def:SequenceIndex}
\end{definition}

Consider the dataset given in Fig~.\ref{Fig:1}. Examples of possible significant sequential patterns are $\overrightarrow{s_1} = A \rightarrow B \rightarrow C \rightarrow D$, $\overrightarrow{s_2} = B \rightarrow C \rightarrow D$, $\overrightarrow{s_3} = C \rightarrow D$. As an example let us consider sequence $\overrightarrow{s_1}$. One may notice that density of instances of type $B$ is significant in the neighborhood spaces created for instances of type $A$. That is 1-length sequence $\overrightarrow{s_1} = A$ will be expanded to  $\overrightarrow{s_1} = A \rightarrow B$ and as the tail event set of $\overrightarrow{s_1}$ the set of instances of type $B$ contained in $V_{N(a1)}$ or $V_{N(a2)}$ or $V_{N(a3)}$ will be remembered. Based on the actual tailEventSet($\overrightarrow{s_1}$), $\overrightarrow{s_1}$ will be expanded with event type $C$ and then, in the next step, with $D$.

The sketch of the ST-Miner algorithm provided in \cite{ref1284:Huang2008} is as follows. First, for each event type in a dataset $F$, a 1-length sequence is created. Then, in a depth-first manner, each sequence is expanded with any event type in $F$, if the value of density ratio between the last event type in the sequence, its tail event set and already considered event type is greater that predefined threshold. If the value is below threshold then sequence is not expanded any more.

\section{Efficient discovering of top-K patterns}
\label{Sec:TopKPatterns}

The problem of discovering top-K patterns in data mining tasks is widely known in literature. \cite{ref1284:Tzvetkov2003} and \cite{ref1284:Han2002} consider the problem of discovering top-K closed sequential patterns in transaction databases with minimal lengths given by parameter \textit{min\_len}. For a given sequential pattern $\overrightarrow{s}$ we say that its length is the number of event types participating in $\overrightarrow{s}$ (f.e. $length(\overrightarrow{s_1}) = 4$).

\begin{definition}{Top-K sequential pattern.}
	We say that a pattern $\overrightarrow{s}$ of minimal length \textit{min\_len} is the K-th pattern, if there are K-1 sequential patterns with minimal length \textit{min\_len} and the sequence index value of each is greater or equal to $SeqIndex(\overrightarrow{s})$.  
	\label{Def:TopKPattern}
\end{definition}

Definitions \ref{Def:SequenceIndex} and \ref{Def:TopKPattern} provide means for formulating algorithm discovering top-K patterns. Informally the approach is as follows: starting with 1-length sequences (that is sequences containing singular event types) expand each sequence in a depth-first manner up to the moment when its length will be at least \textit{min\_len}. We start discovering sequences with the basic value of sequence index threshold equal to $1$. At the same time we maintain ranking of top-K already discovered patterns, where the particular rank of a sequence corresponds to its sequence index value. More formally: the i-th rank $r(\overrightarrow{s}) = i$ of a sequence $\overrightarrow{s}$ is calculated as the number of sequences in the ranking with higher sequence indexes than $SeqIndex(\overrightarrow{s})$ plus one. That is, the rank $1$ is associated with the sequence with minimal length $min\_len$ and the highest sequence index value from already discovered sequences with minimal length $min\_len$, the rank $2$ is associated with the sequence with minimal length $min\_len$ and the second highest sequence index value. The three scenarios are possible:

1. If the length of the sequence $\overrightarrow{s}$ is at least \textit{min\_len}, and if there are few than K patterns in the ranking, then $\overrightarrow{s}$ is inserted into the ranking with the rank corresponding to its sequence index value while preserving decreasing order of sequences in the ranking. 

2. If the length of the sequence $\overrightarrow{s}$ is at least \textit{min\_len} and there are K patterns in the ranking, then if the value of sequence index is greater than the value of sequence index of sequence with K-th rank, then K-th sequence is deleted from the ranking and $\overrightarrow{s}$ is inserted into the ranking on the position corresponding to its sequence index value. $\overrightarrow{s}$ is then expanded in a depth-first manner.   

3.  If the length of the sequence $\overrightarrow{s}$ is at least \textit{min\_len} and there are K patterns in the ranking, then if the value of sequence index is smaller than the value of sequence index of sequence with K-th rank, then $\overrightarrow{s}$ is neither inserted into the ranking nor expanded any more.

\begin{figure}[h!t]
	\centering 
	\includegraphics[width=1\linewidth]{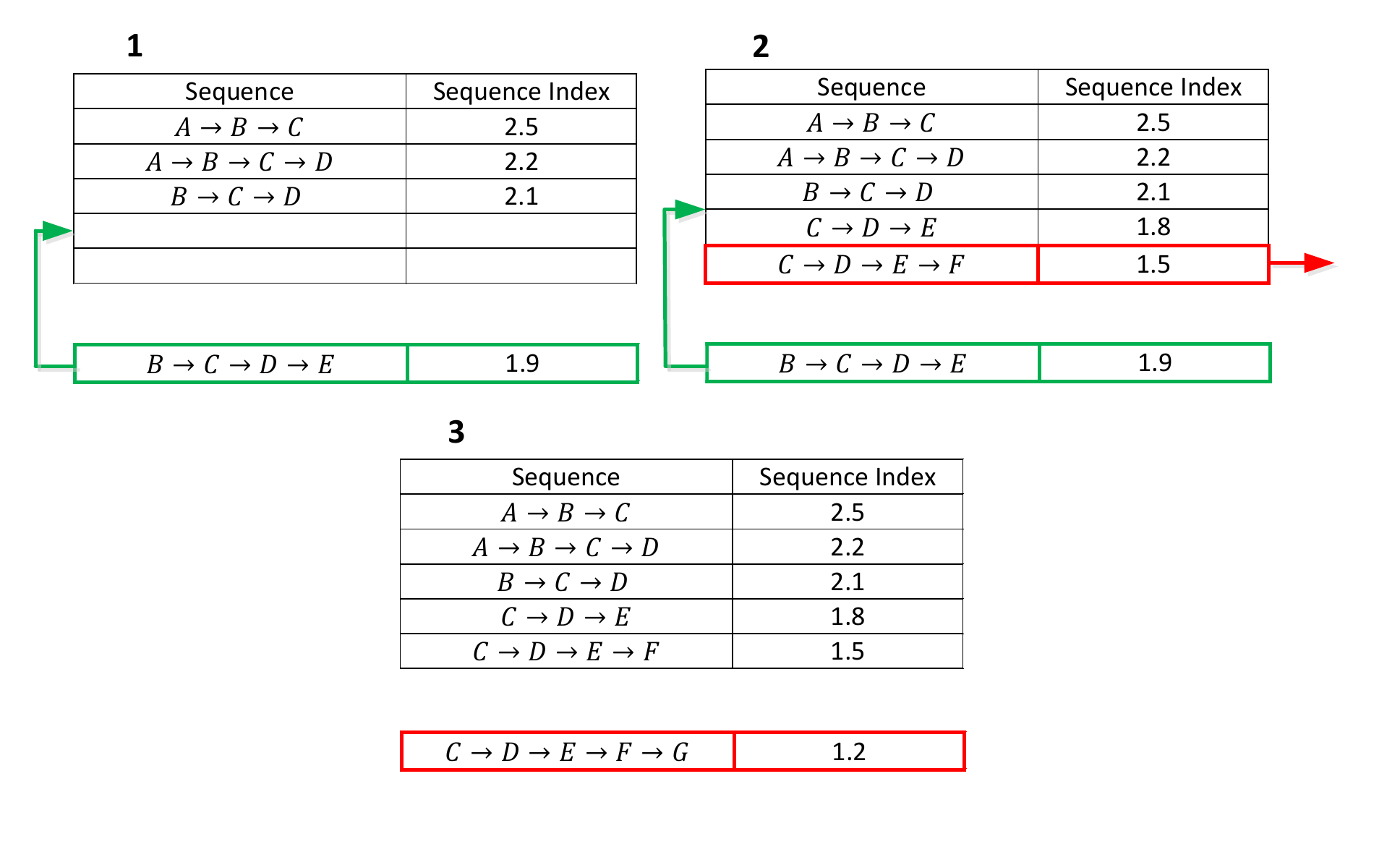}
	\caption{Three possibilities considered inserting a new pattern to the top-K ranking with parameters $min\_len = 3$ and $K = 5$.}
	\label{Fig:2}      
\end{figure}

The above described procedure is shown in Algorithms \ref{Alg:Algorithm1}, \ref{Alg:Algorithm2}, \ref{Alg:Algorithm3}. By $D(f)$ we denote the set of instances of type $f$ in $D$. In Fig. \ref{Fig:2} we show the three above scenarios occurring during considering a pattern to be in the top-K ranking.

\begin{algorithm}[h!t]
	\caption{Procedure for discovering top-K sequential patterns.}
	\begin{algorithmic}[1]
		\Require $D$ - a dataset containing event types and their instances, $ K $ - number of top patterns to discover, $ min\_len $ - minimal length of discovered patterns.
		\Ensure A set of top-K sequential patterns.
		\For {Each event type $ f $}
		\State Create 1-length sequence $ \overrightarrow{s} $ from $ f $.
		\State TailEventSet($ \overrightarrow{s} $) $ \leftarrow $ $ D(f) $.
		\State \textit{ExpandSequence($ \overrightarrow{s} $, $ min\_len $)}.
		\EndFor 
	\end{algorithmic} 
	\label{Alg:Algorithm1}
\end{algorithm}

\begin{algorithm}[h!t]
	\caption{ExpandSequence($\protect \overrightarrow{s}$, $ min\_len $) procedure}
	\begin{algorithmic}[1]
		\Require $ \overrightarrow{s} $ - sequence to be expanded, $ min\_len $ - minimal length of discovered patterns.
		\For {Each event type $f$}
		\State TailEventSet($\overrightarrow{s} \rightarrow f$) $\leftarrow$ SpatialJoin(TailEventSet($\overrightarrow{s}$), $ D(f) $).
		\State Calculate SequenceIndex($ \overrightarrow{s} \rightarrow f $).
		\If {SeqIndex($\overrightarrow{s} \rightarrow f$) $ > 1$ }
		\If {Length($\overrightarrow{s} \rightarrow f$) $\geq min\_len$}
		\If {Number or discovered patterns $<$ K}
		\State Insert $\overrightarrow{s}$ into top-K patterns preserving order of ranks.
		\State \textit{ExpandSequence($ \overrightarrow{s} \rightarrow f $, $min\_len$)}.
		\ElsIf {SeqIndex($\overrightarrow{s} \rightarrow f$) $>$ Sequence index of $K$-th pattern }
		\State Delete K-th pattern from top-K.
		\State Insert $\overrightarrow{s}$ into top-K patterns preserving order of ranks.
		\State \textit{ExpandSequence($ \overrightarrow{s} \rightarrow f $, $min\_len$)}.
		\EndIf
		\Else	
		\State \textit{ExpandSequence($ \overrightarrow{s} \rightarrow f $, $min\_len$)}.				
		\EndIf 
		\EndIf
		\EndFor 
	\end{algorithmic}
	\label{Alg:Algorithm2}
\end{algorithm}

\begin{algorithm}[h!t]
	\caption{Calculate SequenceIndex($\protect \overrightarrow{s} \rightarrow f$) function}
	\begin{algorithmic}[1]
		\Require $ \overrightarrow{s} \rightarrow f$ - a sequence of event types; $\overrightarrow{s}[n]$ - the last event type participating in $\overrightarrow{s}$. 
		\State return min(SeqIndex($\overrightarrow{s}$), DensityRatio($ \overrightarrow{s}[n] \rightarrow f $)).
	\end{algorithmic}
	\label{Alg:Algorithm3}
\end{algorithm}

We have do discuss some statements occurring in Algorithms \ref{Alg:Algorithm1}, \ref{Alg:Algorithm2}, \ref{Alg:Algorithm3}. By $\overrightarrow{s} \rightarrow f$ we denote the fact that $\overrightarrow{s}$ is expanded with an event type $f$. In Algorithm \ref{Alg:Algorithm2}, Spatial Join procedure performed in step 2 calculates a join set between tail event set of $\overrightarrow{s}$ and set of instances $D(f)$. Spatial join may be performed using the \textit{plane sweep} algorithm proposed in \cite{ref1284:Arge1998}.

\section{Experiments}
\label{sec:Exp} 

We conducted experiments on generated datasets. We use the similar generator and notation of dataset names as proposed in \cite{ref1284:Huang2008}. In Table \ref{tab8}, we show computation times for datasets generated with different maximal lengths of a pattern (other proper but not maximal patterns are subsequeces of maximal patterns). Future research in presented topic should focus on verifying proposed approach using real datasets (such as presented in \cite{ref1284:Mohan2012}, \cite{ref1284:Huang2008}). In our experiments we use cubic neighborhood spatiotemporal spaces $V_{N(e)}$ with parameters $\mathcal{R} = 10$ (size of spatial dimension) and $\mathcal{T} = 10$ (size of temporal window). The whole spatiotemporal space $V$ is given by parameters $DSize = 1000$ and $TSize = 1200$.

\begin{table}[h!t]
	\caption{Average computation times (in seconds) for generated datasets using different values K and $min\_len = 3$.}
	\begin{tabularx}{\textwidth}{c>{\centering}cXXXXXXXX}
		\toprule
		\multicolumn{10}{c}{Pn = 4, Ni = 10, Nf = 25, $\mathcal{R} = 10$, $\mathcal{T} = 10$, $DSize = 1000$, $TSize = 1200$} \\
		\midrule \midrule 
		Ps & Avg.~dataset size & \multicolumn{8}{c}{K} \\
		\midrule \midrule 
		& & 20 & 30 & 40 & 50 & 60 & 70 & 80 & 90 \\
		\cmidrule{3-10} 
		\addlinespace 
		2  & 2574 & 1.30 & 1.68 & 2.05 & 2.39 & 2.55 & 2.90 & 3.45 & 4.26 \\ 
		\cmidrule{1-2} \cmidrule{3-10} \addlinespace 
		3  & 6876 & 4.64 & 5.99 & 7.13 & 7.70 & 8.73 & 10.25 & 11.48 & 13.25 \\
		\cmidrule{1-2} \cmidrule{3-10} \addlinespace 
		4 & 10980 & 14.92 & 21.46 & 26.26 & 29.71 & 33.63 & 37.32 & 42.48 & 48.00	\\	
		\cmidrule{1-2} \cmidrule{3-10} \addlinespace 
		5 & 14125 & 12.15 & 16.00 & 20.96 & 24.47 & 24.47 & 26.73 & 31.40 & 34.66 \\
		\cmidrule{1-2} \cmidrule{3-10} \addlinespace 
		6 & 18368 & 10.52 & 27.37 & 30.72 & 35.66 & 39.87 & 44.05 & 49.33 & 55.98 \\
		\bottomrule
	\end{tabularx}
	\label{tab8}
\end{table}

\section{Remarks and conclusion}
\label{Sec:Conclusions}

In the paper, we consider the problem of effective discovering of top-K patterns in event-based spatio-temporal data. In particular, we propose the method creating ranking of top-K already discovered patterns and dynamically updating ranking based on the rank of already expanded pattern. The approach allows to immediately prune patterns which for sure will not be among the top-K patterns with length defined by \textit{min\_len} parameter. In the experiments we show the efficiency of proposed approach.

Future research should focus on investigating properties of described notion of sequential patterns and proposing methods discovering top-K patterns in limited memory environments. Additionally, proposed approach should be verified real data.

\end{document}